\documentstyle[12pt]{article}

\advance\textheight by \topskip
\newcommand{\Dir}{\kern -6.4pt\Big{/}}%su lettere italiane minuscole
\newcommand{\Dirin}{\kern -10.4pt\Big{/}\kern 4.4pt}
\newcommand{\DDir}{\kern -7.6pt\Big{/}}%su lettere italiane maiuscole
\newcommand{\DGir}{\kern -6.0pt\Big{/}}%su lettere greche

\begin{document}

\thispagestyle{empty}
\setcounter{page}{0}

\begin{flushright}
{\large DFTT 68/93}\\
{\rm November 1993\hspace*{.5 truecm}}\\
\end{flushright}

\vspace*{\fill}

\begin{center}
{\Large \bf Five Jet Production with heavy quarks
at $e^+ e^-$ Colliders \footnote{ Work supported in part by Ministero
dell' Universit\`a e della Ricerca Scientifica. \hfill\break\indent
\ e--mail: ballestrero,maina@to.infn.it}}\\[2cm]
{\large Alessandro Ballestrero and Ezio Maina}\\[.3 cm]
{\it Dipartimento di Fisica Teorica, Universit\`a di Torino}\\
{\it and INFN, Sezione di Torino}\\
{\it v. Giuria 1, 10125 Torino, Italy.}\\
\end{center}

\vspace*{\fill}

\begin{abstract}
{\normalsize
Heavy quark production in five jet events at $e^+e^-$ colliders
is studied at tree level using helicity amplitudes. Total production rates
for $2b3j$ and $4bj$ are given and compared with the corresponding results
for massless quarks. The process $e^+e^-\rightarrow q\bar q gg\gamma$ which
is the dominant contribution to $4j\gamma$ production is briefly discussed.
}
\end{abstract}

\vspace*{\fill}

\newpage
\subsection*{Introduction}
The great number of hadronic decays of the $Z^0$ observed at LEP
provides the opportunity to test our understanding of strong interactions
in unprecedented detail.
Large samples of multi--jet events
have been accumulated and analyzed \cite{expMC,alphas}.
In particular the ALEPH Collaboration and the OPAL Collaboration have
studied the jet--fraction distributions as a function of $y_{cut}$ for
up to five jets. \par
Recent advances in $b$--tagging techniques based on the introduction of
vertex detectors and on a refinement of the selection procedures,
with their large efficiencies and the resulting high purities, have paved
the way to the study of heavy--quark production in association with
light--quark and gluon jets. The flavour independence of the strong coupling
constant has been verified by several groups \cite{alphab}
using both jet--rates and shape variables.\par
It has been lately pointed out \cite{last1,last2}
that the effects of the $b$--quark mass are substantial
and increase with the number of jets.\par
In this paper we plan to study heavy--quark production in five--jet events
at $e^+e^-$ colliders.
Two different mechanisms contribute to five--jet production for light quarks,
the `point--like' annihilation of $e^+e^-$ to $Z^0,\gamma$ \cite{fgk},
which dominates at LEP I energy, and $W^+W^-$ production \cite{brown},
followed by
the emission of a gluon from the decay products of the $W$'s, which
dominates above the $WW$ treshold. This latter contribution is severely
suppressed for $b$--quarks by the smallness of the $V_{bc}$ element of
the CKM matrix and by the large mass of the $top$ and will not be
considered here.
We have studied the
annihilation mechanism for five--jet production at tree
level taking full account of $\gamma, Z$ interference and of quark masses.
Therefore our matrix elements can be used  at all $e^+e^-$ colliders.\par
As a byproduct of the main calculation we have also examined the dominant
contribution, $e^+e^-\rightarrow q\bar q gg\gamma$, to the production of a
single hard photon in association with four jets. Both initial state
and final state radiation have been included at tree level,
without ISR resummation.\par
We have computed all matrix elements both in the formalism of \cite{ks,mana}
and in that of \cite{hz} using the former for the actual calculation.
The relevant formulae are summarized in the following section.
Both methods can be easily implemented in a small set of nested subroutines.
This however results in a computer program which is far too slow for the
present calculation. We have instead used the symbolic package $Mathematica$
\cite{math}
to write down the Fortran expression for each helicity amplitude.
The individual $Z$-functions are computed first and then combined in
larger structures that describe chunks of
increasing size of the Feynman diagrams shown in fig.1--2.
Every sub--diagram is saved and then used several times.
With this procedure we have produced a rather large piece of code,
which however runs quite fast, and therefeore
can be used in high statistics Montecarlo runs.
As an example the program for $q\bar q ggg$ production is about
24,000 lines long, but requires only about $5\times 10^{-2}$ seconds
to evaluate on a Vaxstation 4000/90.
\par
The amplitudes have been checked for gauge invariance.
We have used  $M_Z=91.1$ GeV, $\Gamma_Z=2.5$ GeV,
$\sin^2 (\theta_W)=.23$, $m_b=5.$ GeV, $\alpha_{em}= 1/128$ and
$\alpha_{s}= .115$ in the numerical part of our work.\par

\subsection*{Calculation}
We present the helicity amplitudes in the formalism of \cite{ks,mana}.
The two basic functions which are needed in writing the amplitudes are:
\begin{eqnarray}\label{zfunc}
\lefteqn{ \hspace{-.5in}
         Z(p_{i},\lambda_{i};p_{j},\lambda_{j};p_{k},
                   \lambda_{k};p_{l},\lambda_{l};
           c_R,c_L;c'_R,c'_L) =}\hspace{.5in}\\ \nonumber
& &[\bar u(p_{i},\lambda_{i}) \Gamma^{\mu} u(p_{j},\lambda_{j})]
[\bar u(p_{k},\lambda_{k}) \Gamma'_{\mu} u(p_{l},\lambda_{l})]
\end{eqnarray}
where
\begin{equation}
\Gamma^{(')\mu}=\gamma^{\mu}
(c^{(')}_R P_R + c^{(')}_L P_L)
\end{equation}
and
\begin{equation}\label{wfunc}
X(p_{i},\lambda_{i};p_{j};p_{k},\lambda_{k}) =
[\bar u(p_{i},\lambda_{i}) {p_{j}}\Dirin u(p_{k},\lambda_{k})]
\end{equation}
Their expressions in terms of spinor products
$s(i,j) = \bar u(p_{i},+) u(p_{j},- ) = t^\ast(j,i)$,
$\eta$ and $\mu$ functions
can be found in Table I and II.\par
A convenient, though non unique, choice for the spinor products
and for the functions $\eta$ and $\mu$ is:
\begin{equation}\label{st}
s(i,j)  =
(p^y_i + i p^z_i) \left( {{p^0_j - p^x_j} \over {p^0_i - p^x_i}}
\right)^{1/2} -
\;(p^y_j + i p^z_j) \left( {{p^0_i - p^x_i} \over {p^0_j - p^x_j}}
\right)^{1/2}
\end{equation}
\begin{equation}\label{etamu}
\eta(i)  =  \left( 2( p^0_i - p^x_i) \right)^{1/2} \kern .5 in
\mu(i)  =   \pm {{m_i}\over{\eta(i)}}
\end{equation}
where the sign $+(-)$ refers to a particle (antiparticle)
of mass $m_i$.\par
The polarization vectors of a gluon of momentum $p_i$ can be
written as
\begin{equation}\label{eps}
\varepsilon^\mu(p_i,\lambda) = N_i
[\bar u(p_i,\lambda)\gamma^\mu u(p_{a(i)},\lambda)]
\end{equation}
where $p_{a(i)}$ is an auxiliary massless
momentum  not parallel to $p_i$ and the normalization factor is
$N_i = [4(p_{a(i)}\cdot p_i)]^{-1/2}$.
All amplitudes are independent of the $p_{a(i)}$'s.\par
There are fifty Feynman diagrams contributing to
\begin{equation}\label{pro1}
e^-(p_1,\lambda)+e^+(p_2,-\lambda)
\rightarrow
q(p_6,\lambda_6)+\bar q(p_7,\lambda_7)+
g(p_3,\lambda_3)+ g(p_4,\lambda_4)+ g(p_5,\lambda_5)
\end{equation}
Some of them are shown in fig.1.
All others can be obtained through permutations of the gluon labels.
The two diagrams with four-gluon vertices are not shown. Their
contribution
is included in the expressions given below for the
diagrams with two connected
three--gluon vertices.\par
The amplitude squared for this process is:
\begin{equation}\label{msquare1}
{\left|{\overline M}\right|}_{q\bar q ggg}^2=
  {1\over 4} {1\over {3!}} g_s^3 e^2 N_3 N_4 N_5 \sum_{\{\lambda\}}
\sum_{\pi, \pi^\prime}{M}_\pi^{\{\lambda\}}
M_{\pi^\prime}^{\{\lambda\}*} C_{\pi\pi^\prime}
\end{equation}
where $\pi$ is a permutation of the indexes $(3,4,5)$,
$\{\lambda\}$ indicates the helicities of all external particles,
$M_{\pi}$ is the coefficient of the color matrix
\begin{equation}\label{tprod}
T^{\pi}_{ij} = \left( T^{\pi(3)}T^{\pi(4)}T^{\pi(5)} \right)_{ij}
\end{equation}
in the full amplitude and $C_{\pi\pi^\prime}$ is the appropriate
color factor,
$C_{\pi\pi^\prime} = {\rm Tr}(T^\pi (T^{\pi^\prime})^\dagger)$.
\par
The color factors are given in Table III. To fix our notation, diagrams
1--4 have color structure 345, diagrams 5--7 have color structure [3,4]5
where [3,4] is the usual commutator, [3,4]=34$-$43,
diagrams 8--10 have color structure 5[3,4] and
diagrams 11--12 have color structure [[3,4],5].\par
The production of four quarks and one gluon
\begin{equation}\label{pro2}
e^-(p_1,\lambda)+e^+(p_2,-\lambda)
\rightarrow
g(p_3,\lambda_3) + q(p_4,\lambda_4)+\bar q(p_5,\lambda_5)+
q(p_6,\lambda_6)+ \bar q(p_7,\lambda_7)
\end{equation}
is described in lowest order by twenty--four Feynman diagrams
if the two quarks have
different flavour (case I) and by forty--eight if their
flavour is the same
(case II).
Twelve of the diagrams for case I are shown in fig.2, all others
can be obtained with the simultaneous substitutions
$\bar q(p_5,\lambda_5)\leftrightarrow \bar q(p_7,\lambda_7)$ and
$ q(p_4,\lambda_4)\leftrightarrow q(p_6,\lambda_6)$.
The additional ones needed for identical flavour can be obtained
exchanging $\bar q(p_5,\lambda_5)$ with $\bar q(p_7,\lambda_7)$
including their color quantum numbers and changing the overall sign.
\par
The amplitude squared for these processes is:
\begin{equation}\label{msquare2}
{\left|{\overline M}\right|}_{q\bar q q^\prime\bar q^\prime g}^2=
  {1\over 4} g_s^3 e^2 N_3 \sum_{\{\lambda\}}
\sum_{i,j=1}^{24}{M}_i^{\{\lambda\}}
M_j^{\{\lambda\}*} C_{ij}
\end{equation}
\begin{equation}\label{msquare3}
{\left|{\overline M}\right|}_{q\bar q q\bar q g}^2=
  {1\over 4} {1\over {(2!)^2}} g_s^3 e^2 N_3 \sum_{\{\lambda\}}
\sum_{i,j=1}^{48}{M}_i^{\{\lambda\}}
M_j^{\{\lambda\}*} C_{ij}^\prime
\end{equation}
in case I and case II respectively.
$C_{ij}$ and $C_{ij}^\prime$ are the appropriate
colour factors which can be extracted from Table IV and V.\par
We define the following colour structures:
\begin{eqnarray}\label{colour}
A_{1\ i_1j_1}^{a\ i_2j_2} &=& \left( T^aT^b\right)_{i_1j_1}
                              \left( T^b\right)_{i_2j_2} \nonumber \\
A_{2\ i_1j_1}^{a\ i_2j_2} &=& \left( T^bT^a\right)_{i_1j_1}
                              \left( T^b\right)_{i_2j_2} \nonumber \\
A_{3\ i_1j_1}^{a\ i_2j_2} &=& \left( T^b\right)_{i_1j_1}
                              \left( T^aT^b\right)_{i_2j_2} \\
A_{4\ i_1j_1}^{a\ i_2j_2} &=& \left( T^b\right)_{i_1j_1}
                              \left( T^bT^a\right)_{i_2j_2} \nonumber \\
A_{5\ i_1j_1}^{a\ i_2j_2} &=& \left( T^b\right)_{i_1j_1}
                              \left( T^c\right)_{i_2j_2}
c_{abc}\nonumber \\
\end{eqnarray}
and
\begin{equation}
B_{n\ i_1j_1}^{a\ i_2j_2} = A_{n\ i_1j_2}^{a\ i_2j_1}.
\end{equation}
The color factor is given by:
\begin{equation}
C_{nm} = A_{n\ i_1j_1}^{a\ i_2j_2} {A^\dagger}_{m\ j_1i_1}^{a\ j_2i_2}
\end{equation}
in case I and by analogous formulae in case II, possibly with
$A_{n,m} \rightarrow B_{n,m}$.\par
We adopt the following shorthand notation,
where in the left hand side the index $i$ stands both for the momentum
$p_{i}$ and for the corresponding polarization $\lambda_i$,
and $[i]$ stands for the set
of four indices associated with a gluon momentum and to
his auxiliary momentum:
\begin{equation}\label{xshort}
X(i;j;k) =
X(p_{i},\lambda_{i};p_{j};p_{k},\lambda_{k})
\end{equation}
\begin{equation}\label{xglue}
X([i];j) = X(p_{i},\lambda_{i};p_{j};p_{a(i)},\lambda_{i})
\end{equation}
\begin{equation}\label{zshort}
Z(i,j;k,l) = Z(p_{i},\lambda_{i};p_{j},\lambda_{j};p_{k},
\lambda_{k};p_{l},\lambda_{l};1,1;1,1)
\end{equation}
\begin{equation}\label{zglue}
Z([i];j,k) = Z(p_{i},\lambda_{i};p_{a(i)},\lambda_{i};p_{j},\lambda_{j};
p_{k},\lambda_{k};1,1;1,1)
\end{equation}
\begin{equation}\label{zglueglue}
Z([i];[j]) = Z(p_{i},\lambda_{i};p_{a(i)},\lambda_{i};
p_{j},\lambda_{j};p_{a(j)},\lambda_{j};1,1;1,1)
\end{equation}
\begin{eqnarray}\label{zgamma}
Z_e(i,j) & = & Z(p_{i},\lambda_{i};p_{j},\lambda_{j};
p_{2},\lambda;p_{1},\lambda;1,1;1,1)
\times d_\gamma \\ \nonumber
& & \mbox{} + Z(i,\lambda_{i};j,\lambda_{j};2,\lambda;1,\lambda;
c_R,c_L;c'_R,c'_L) \times d_{Z^0}
\end{eqnarray}
where $d_\gamma$, $d_{Z^0}$ are the photon and $Z^0$ propagator functions.
The attachment of two external gluons to a quark line through a
three--gluon vertex is conveniently described with the help of:
\begin{eqnarray}\label{z2}
\hspace{-.4in}
Z_2([i,j];k,l) & = & Z([i];[j])
\times \left( \rule [-.1cm]{0cm}{.6cm}
          X(k;j;l) - X(k;i;l) \right) \\
& & \mbox{} + 2 \left( \rule [-.1cm]{0cm}{.6cm} X([j];i)
                  \times Z([i];k,l)
                      - X([i];j)
                  \times Z([j];k,l) \right) \nonumber
\end{eqnarray}
The function that describes the insertion of two connected three--gluon
vertices, and of
the term with the same color structure in the
associated four--gluon vertex,
on a fermion line is:
\begin{eqnarray}\label{z3}
\lefteqn{Z_3([i,j],[k];l,m) =} \nonumber \\
&  & 4 Z([i];l,m)
\times X([j];i) \times
        \left( \rule [-.1cm]{0cm}{.6cm}
X([k];j) + X([k];i)  \right) \nonumber \\
& &  \mbox{}+ 2 Z([k];l,m) \times \left( \rule [-.2cm]{0cm}{.7cm}
              Z([i];[j])
\times p_{i}\cdot p_{k} \right. \nonumber \\
& & \hspace{.4 in}\left.
         +X([i];j)\times \left (X([j];i)+ 2 X([j];k)
            \rule [-.1cm]{0cm}{.6cm} \right)
\rule [-.2cm]{0cm}{.7cm}\right) \\
& & \mbox{}+\left( \rule [-.1cm]{0cm}{.6cm}
             X(l;k;m) - X(l;i;m) - X(l;j;m)\right)
                   \times \nonumber \\
& & \hspace{.4 in}
     \left( \rule [-.1cm]{0cm}{.6cm}
               Z([i];[j])\times X([k];j) - 2 Z([k];[j])
                \times X([i];j)\right)
                     \nonumber \\
& & \mbox{}+ 2 X(l;j;m)\times
                   Z([i];[j])\times
        \left( \rule [-.1cm]{0cm}{.6cm} X([k];j) + X([k];i)\right)
                      \nonumber \\
& & \mbox{}- 2 Z([i];l,m)\times Z([k];[j])\times p_{i}\cdot p_{j}
                      \nonumber \\
& & \hspace{1.2 in} - ( i \leftrightarrow j )
\nonumber
\end{eqnarray}
The denominator associated with this function is $(p_{i}+p_{j})^2
(p_{i}+p_{j}+p_{k})^2$.\par
In order to simplify our formulae we make the convention
that all repeated indices are summed over, and that the sum extends to
all possible values, $\lambda = \pm 1$, of all internal helicities.
The last line in eq. \ref{m1}--\ref{m24} gives the set of values taken
independently by each momentum index.
Defining the sign factors $\varepsilon(i)$ as:
\begin{eqnarray}\label{sign}
\varepsilon(i) = \left\{ {\begin{array}{ll} -1 & i=1,2  \\
                                            +1 & i > 2
                          \end{array}}
                 \right.
\end{eqnarray}
the spinor functions necessary for $q\bar q ggg$ production,
which correspond to the diagrams shown in fig.1 are:
\begin{eqnarray}\label{m1}
{M}_1 & = & -
Z([3];6,i)\times Z([4];i,j)\times Z([5];j,k)
\times Z_e(k,7)\times \varepsilon(k) \\
      &   & p_{i}=\{p_{6},p_{3}\}, \enskip
     p_{j}=\{p_{6},p_{3},p_{4}\}, \enskip p_{k}=\{p_{1},p_{2},p_{7}\}
           \nonumber
\end{eqnarray}
\begin{eqnarray}\label{m2}
{M}_2 & = & -
Z([3];6,i)\times Z([4];i,j)\times Z_e(j,k)
\times Z([5];k,7) \\
      &   & p_{i}=\{p_6,p_3\}, \enskip p_{j}=\{p_6,p_3,p_4\},
 \enskip p_{k}=\{p_5,p_7\}
           \nonumber
\end{eqnarray}
\begin{eqnarray}\label{m3}
{M}_3 & = &
Z([3];6,i)\times Z_e(i,j)\times
Z([4];j,k)\times Z([5];k,7) \\
      &   &p_{ i}=\{p_6,p_3\}, \enskip p_{j}=\{p_4,p_5,p_7\},
              \enskip p_{k}=\{p_5,p_7\}
           \nonumber
\end{eqnarray}
\begin{eqnarray}\label{m4}
{M}_4 & = &
Z_e(6,i) \times Z([3];i,j)\times
Z([4];j,k)\times Z([5];k,7)\times\varepsilon(i) \\
      &   & p_i=\{p_6,p_1,p_2\}, \enskip p_j=\{p_4,p_5,p_7\},
       \enskip p_k=\{p_5,p_7\}
           \nonumber
\end{eqnarray}
\begin{eqnarray}\label{m5}
{M}_{5} & = &
-Z_2([3,4];6,i) \times Z([5];i,j)
\times Z_e(j,7)\times\varepsilon(j)\\
         &   & p_i=\{p_6,p_3,p_4\}, \enskip p_j=\{p_1,p_2,p_7\}
           \nonumber
\end{eqnarray}
\begin{eqnarray}\label{m6}
{M}_{6} & = &
-Z_2([3,4];6,i) \times Z_e(i,j)
\times Z([5];j,7)\\
         &   & p_i=\{p_6,p_3,p_4\}, \enskip p_j=\{p_5,p_7\}
           \nonumber
\end{eqnarray}
\begin{eqnarray}\label{m7}
{M}_{7} & = &
-Z_e(6,i) \times Z_2([3,4];i,j)
\times Z([5];j,7)\times\varepsilon(i)\\
         &   & p_i=\{p_1,p_2,p_6\}, \enskip p_j=\{p_5,p_7\}
           \nonumber
\end{eqnarray}
\begin{eqnarray}\label{m8}
{M}_{8} & = & -Z([5];6,i) \times Z_2([3,4];i,j)
\times Z_e(j,7)\times\varepsilon(j)\\
         &   & p_i=\{p_6,p_5\}, \enskip p_j=\{p_1,p_2,p_7\}
           \nonumber
\end{eqnarray}
\begin{eqnarray}\label{m9}
{M}_{9} & = & -Z([5];6,i) \times Z_e(i,j)
\times Z_2([3,4];j,7)\\
         &   & p_i=\{p_6,p_5\}, \enskip p_j=\{p_3,p_4,p_7\}
           \nonumber
\end{eqnarray}
\begin{eqnarray}\label{m10}
{M}_{10} & = &
-Z_e(6,i) \times Z([5];i,j)\times Z_2([3,4];j,7)
\times\varepsilon(i)\\
         &   & p_i=\{p_1,p_2,p_6\}, \enskip p_j=\{p_3,p_4,p_7\}
           \nonumber
\end{eqnarray}
\begin{eqnarray}\label{m11}
{M}_{11} & = &
- Z_3([3,4],[5];6,i) \times Z_e(i,7)
\times\varepsilon(i)\\
         &   & p_i=\{p_1,p_2,p_7\}
           \nonumber
\end{eqnarray}
\begin{eqnarray}\label{m12}
{M}_{12} & = &
Z_e(6,i) \times Z_3([3,4],[5];i,7)
\times\varepsilon(i)\\
         &   & p_i=\{p_1,p_2,p_6\}\nonumber
\end{eqnarray}
The spinor functions for the diagrams in fig.2, which describe
$q\bar q q\bar q g$ production are:
\begin{eqnarray}\label{m13}
{M}_{13} & = &
Z([3];i,5)\times Z(4,i;6,j)\times Z_e(j,7)
              \times\varepsilon(j) \\
      &   & p_i=\{p_3,p_5\}, \enskip p_j=\{p_1,p_2,p_7\} \nonumber
\end{eqnarray}
\begin{eqnarray}\label{m14}
{M}_{14} & = &
- Z([3];4,i)\times Z(i,5;6,j)\times Z_e(j,7)
              \times\varepsilon(j) \\
      &   & p_i=\{p_3,p_4\}, \enskip p_j=\{p_1,p_2,p_7\} \nonumber
\end{eqnarray}
\begin{eqnarray}\label{m15}
{M}_{15} & = &
- Z([3];i,5)\times Z(4,i;j,7)\times Z_e(6,j)
              \times\varepsilon(j) \\
      &   & p_i=\{p_3,p_5\}, \enskip p_j=\{p_1,p_2,p_6\} \nonumber
\end{eqnarray}
\begin{eqnarray}\label{m16}
{M}_{16} & = &
Z([3];4,i)\times Z(i,5;j,7)\times Z_e(6,j)
              \times\varepsilon(j) \\
      &   & p_i=\{p_3,p_4\}, \enskip p_j=\{p_1,p_2,p_6\} \nonumber
\end{eqnarray}
\begin{eqnarray}\label{m17}
{M}_{17} & = &
-Z(6,i;4,5)\times Z([3];i,j)\times  Z_e(j,7)
              \times\varepsilon(j) \\
  &   & p_i=\{p_4,p_5,p_6\}, \enskip p_j=\{p_1,p_2,p_7\} \nonumber
\end{eqnarray}
\begin{eqnarray}\label{m18}
{M}_{18} & = &
- Z(6,i;4,5)\times Z_e(i,j) \times  Z([3];j,7)
              \\
      &   & p_i=\{p_4,p_5,p_6\}, \enskip p_j=\{p_3,p_7\} \nonumber
\end{eqnarray}
\begin{eqnarray}\label{m19}
{M}_{19} & = &
- Z_e(6,i)\times  Z(i,j;4,5)\times Z([3];j,7)
              \times\varepsilon(i) \\
      &   & p_i=\{p_1,p_2,p_6\}, \enskip p_j=\{p_3,p_7\} \nonumber
\end{eqnarray}
\begin{eqnarray}\label{m20}
{M}_{20} & = &
-Z([3];6,i)\times Z(i,j;4,5)\times   Z_e(j,7)
              \times\varepsilon(j) \\
      &   & p_i=\{p_3,p_6\}, \enskip p_j=\{p_1,p_2,p_7\} \nonumber
\end{eqnarray}
\begin{eqnarray}\label{m21}
{M}_{21} & = &
- Z([3];6,i) \times   Z_e(i,j) \times Z(j,7;4,5)
              \\
      &   & p_i=\{p_3,p_6\}, \enskip p_j=\{p_4,p_5,p_7\} \nonumber
\end{eqnarray}
\begin{eqnarray}\label{m22}
{M}_{22} & = &
-Z_e(6,i) \times Z([3];i,j) \times  Z(j,7;4,5)
              \times\varepsilon(i) \\
  &   & p_i=\{p_1,p_2,p_6\}, \enskip p_j=\{p_4,p_5,p_7\} \nonumber
\end{eqnarray}
\begin{eqnarray}\label{m23}
{M}_{23} & = &
Z_e(6,i) \times \left(\rule[-.2cm]{0cm}{.7cm}
                2\left(\rule[-.1cm]{0cm}{.6cm}
                  X([3];4)+X([3];5)\right)\times Z(i,7;4,5)
                          \right. \nonumber \\
      &   &  \hspace{.2 in}     -2 X(4;3;5)\times Z([3];i,7)\\
      &   &  \left. -\left( \rule [-.1cm]{0cm}{.6cm}
           X(i;4;7)+X(i;5;7)-X(i;3;7)\right)\times Z([3];4,5)
                   \rule [-.2cm]{0cm}{.7cm} \right)
              \times\varepsilon(i) \nonumber \\
      &   & \hspace{.5 in} p_i=\{p_1,p_2,p_6\} \nonumber
\end{eqnarray}
\begin{eqnarray}\label{m24}
{M}_{24} & = &
-Z_e(i,7) \times \left( \rule [-.1cm]{0cm}{.6cm}
                  2\left(\rule [-.1cm]{0cm}{.6cm}
               X([3];4)+X([3];5)\right)\times Z(6,i;4,5)
                           \right. \nonumber \\
      &   & \hspace{.2 in}  -2 X(4;3;5)\times Z([3];6,i)\\
      &   & \left. -\left( \rule [-.1cm]{0cm}{.6cm}
            X(6;4;i)+X(6;5;i)-X(6;3;i)\right)\times Z([3];4,5)
                       \rule [-.2cm]{0cm}{.7cm}\right)
              \times\varepsilon(i) \nonumber \\
      &   & \hspace{.5 in} p_i=\{p_1,p_2,p_7\} \nonumber
\end{eqnarray}
{}From the examples above the one can to construct the
full amplitudes for $5j$ and $4j\gamma$ production with
little effort.\par
\subsection*{Results}
Our results are presented in fig.3 and 4. In fig.3 we show
the total cross section for five--jet production as a function
of $y_{cut}$ in the
JADE scheme \cite{jade} and in the DURHAM scheme \cite{durham}.
We have neglected the mass of the quarks of the first two
generations, and
have summed over all their contributions. The cross section for four
$b$--quark plus one gluon--jet production is given by the dotted lines.
The dashed lines represent the cross section for the production of
two $b$--quarks plus any combination of three light--parton jets.
Finally the sum of all contributions to five--jet production including
only light, strongly interacting, particles is shown by the continuous line.
It is to be noticed that, at least in the JADE scheme
in which $y$, neglecting the mass of the light hadrons which are
effectively detected, represents an actual invariant mass,
there is a natural lower
limit to the allowed range in $y_{cut}$ for $b$--quark production,
which corresponds to the invariant
mass of the $b$,
$y_{cut}^{\rm min} \approx (m_b/M_{Z^0})^2\approx 3\times 10^{-3}$.
At lower $y_{cut}$ one begins to resolve the decay products of the heavy
quark.\par
In fig.3 we also give, in the JADE scheme, the cross section for
$e^+e^-\rightarrow q\bar q gg\gamma$, summing over five flavours.
In this case we have taken for the $c$--quark $m_c=1.7$ GeV.
Additional cuts are necessary
in order to screen our results from the collinear singularities
due to initial state radiation. We have required $p^\gamma_t > 5$ GeV
and ${\mid\cos\theta_\gamma\mid < .72}$; these cuts are similar
to the selection
criteria adopted by the LEP  collaborations.\par
When our results for five--jet production are compared with the data
presented by ALEPH
\cite{expMC} and OPAL \cite{alphas}
it is clear that the absolute normalization is about a factor of five
too small. The simplest explanation for this discrepancy is
our choice for $\alpha_s$.
In fact we have used $\alpha_s =.115$ which corresponds to
$Q^2= M^2_{z^0}$ with $\Lambda_{\overline{MS}}=200$ MeV with five
active flavours in the standard formula:
\begin{equation}
\alpha_s(Q) = {{1}\over{b_0 \log (Q^2/\Lambda^2)}} \left[
1- {{b_1 \log (\log (Q^2/\Lambda^2))}\over{b_0^2 \log (Q^2/\Lambda^2) }}
\right]
\end{equation}
The analysis of shape variables and jet rates to ${\cal O}(\alpha_s^2)$
has shown that, in order to get agreement between the data and the
theoretical predictions, the scale of the strong coupling constant
has to be chosen to be $Q= x_\mu M_{Z^0}$, with $x_\mu \approx 0.1$
\cite{alphas}.
It has later been shown that when the relevant logarithms are properly
resummed \cite{catani} agreement is obtained for much larger
values of the scale, $x_\mu \approx 1.$ \cite{resummed}.
It therefore not surprising that our tree level expressions require
a relatively small scale in order to describe the data.
A related issue, for processes with massive quarks, is the
choice of the scale
of the running mass. Within our tree--level approach we have i
gnored this problem,
using $m_b = 5.$ GeV.
\par
The cross section for 2$b$3j is quite large, of the order of several
hundreds $pb$ at $y_{cut}=.001$ in the
DURHAM scheme and $y_{cut}=.005$ in the JADE scheme,
and it remains sizable out to $y_{cut}\approx .01$
and to $y_{cut}\approx .02$ respectively.
Over this limited range $\sigma (y_{cut})$ decreases by more than one order
of magnitude. This behaviour should be clearly observable with an
efficiency of
order .3 for detecting each $b$--quark.
It will be more difficult to detect the production of 4$b$j, whose
cross section is only a few $pb$ at the lowest values of $y_{cut}$
considered in this paper.\par
In fig.4 we present the cross section ratios $\sigma(2b3g)/\sigma(2d3g)$
(continuous line), $\sigma(2u2bg)/\sigma(2u2dg)$ (dashed line)
and $\sigma(2d2bg)/\sigma(2d2sg)$ (dotted line) in the two
recombination schemes as a function of $y_{cut}$.
These curves confirm our previous conclusions
that mass effects increase with the number of final state light partons.
The ratio for the dominant $2q3g$ production process is equal to
.58 at $y_{cut}=.001$ in the
DURHAM scheme and to .67 at $y_{cut}=.005$ in the JADE scheme.
It is even smaller for the processes with four quark jets in the final state.
This corresponds to a 6$\div$8\% decrease in the predictions for the total
five--jet cross section.\par
An estimate of the effects due to the $charm$ mass
can be obtained from the ratio
$\sigma(2c2g\gamma)/\sigma(2u2g\gamma)$ which is equal to .93 at
$y_{cut}=.005$ in the JADE scheme. The corresponding value for
$\sigma(2b2g\gamma)/\sigma(2d2g\gamma)$ is .63, slightly smaller than
in the case of $2q3g$ production.\par
Five--jet production will be visible also at higher energies.
At $\sqrt s = 200$ GeV we obtain
$\sigma (e^+e^-\rightarrow d\bar d 3g) = 9.8 \times 10^{-2}$ $pb$ and
$\sigma (e^+e^-\rightarrow b\bar b 3g) = 8.8 \times 10^{-2}$ $pb$,
showing that the mass of the $b$ decreases the cross section by 10\%
even at the highest energy envisaged for LEP II.
\subsection*{Conclusions}
The cross section for the production of $b$--quarks in five--jet events
has been computed using helicity amplitudes and the $Z$--function formalism.
The differences with the corresponding massless rates and the total
cross sections for $2b3j$ and $4bj$ production have been studied at the $Z^0$
peak. Mass effects result in a 30$\div$40\% reduction on individual cross
sections and in a 6$\div$8\% reduction of the total five jet production.
The production of a hard photon in association with four jets has been
investigated.
\subsection*{Acknowledgements}
We gratefully acknowledge the collaboration of S. Moretti
in checking the amplitudes for purely hadronic final states
and the collaboration of M. Ughetti for the calculation of
the processes with a hard photon.

\newpage

\subsection*{Table Captions}

\begin{description}

\item[table I] The Z--functions for all independent helicity combinations
in terms of the functions $s$, $t$, $\eta$ and $\mu$ defined in the text.
The remaining Z--functions can be obatined changing sign to the helicities
and exchanging $+$ with $-$, $s$ with $t$ and $R$ with $L$.

\item[table II] The X--functions for the two independent helicity combinations
in terms of the functions $s$, $t$, $\eta$ and $\mu$ defined in the text.
The remaining X--functions can be obatined changing sign to the helicities
and exchanging $+$ with $-$, $s$ with $t$ and $R$ with $L$.

\item[table III] The color factors, multiplied by 9, for $e^+e^-\rightarrow
q\bar q ggg$.

\item[table IV] The color factors, multiplied by 9, for $e^+e^-\rightarrow
q_1\bar q_1 q_2 \bar q_2$.

\item[table V] The additional color factors, multiplied by 9,
needed for quarks of identical flavour $e^+e^-\rightarrow q\bar q q\bar q$.

\end{description}

\newpage

\subsection*{Figure Captions}

\begin{description}

\item[Fig. 1] Representative diagrams contributing to $e^+e^-\rightarrow
        q\bar q ggg$. All others diagrams can be obtained through
     permutations of the gluon labels. The contribution of the two diagrams
   involving a four--gluon vertex, which are not shown, are inluded
in the analytical expression given in the
text for the diagrams with two connected three--gluon vertices.

\item[Fig. 2] Representative diagrams contributing to $e^+e^-\rightarrow
        q\bar q q \bar q$. The remaining diagrams can be obtained through
  simultaneous substitutions
$\bar q(p_5,\lambda_5)\leftrightarrow \bar q(p_7,\lambda_7)$ and
$ q(p_4,\lambda_4)\leftrightarrow q(p_6,\lambda_6)$.
The additional ones needed for identical flavour can be obtained
exchanging $\bar q(p_5,\lambda_5)$ with $\bar q(p_7,\lambda_7)$
including their color quantum numbers and changing the overall sign.

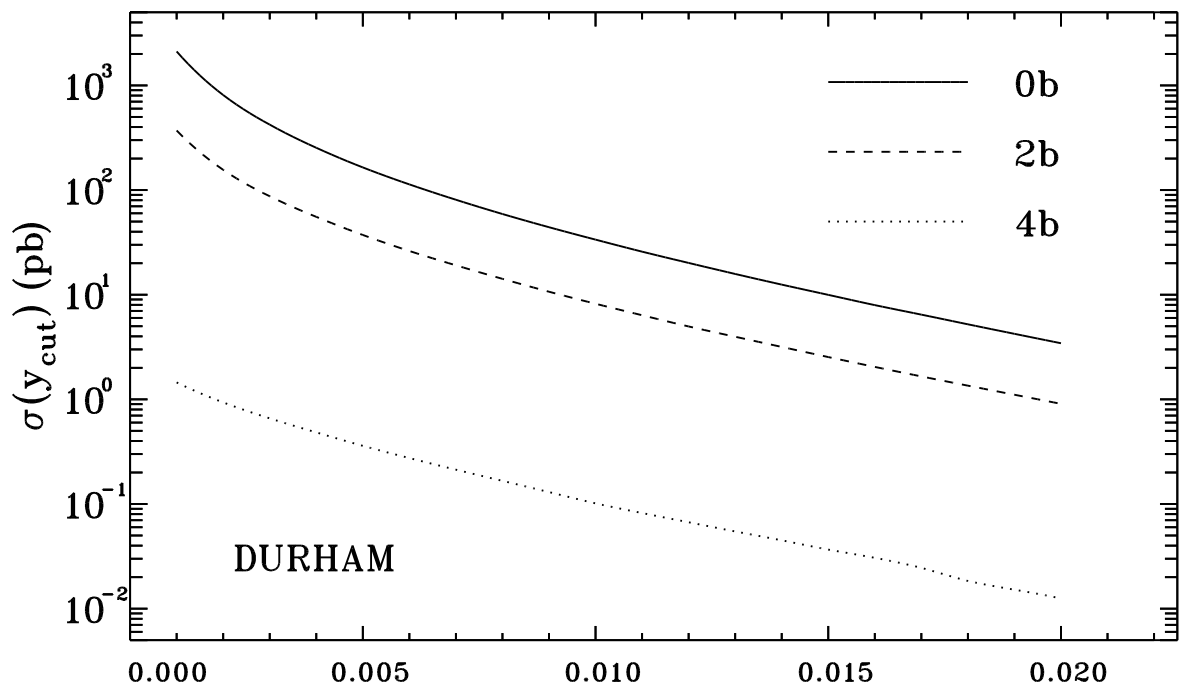
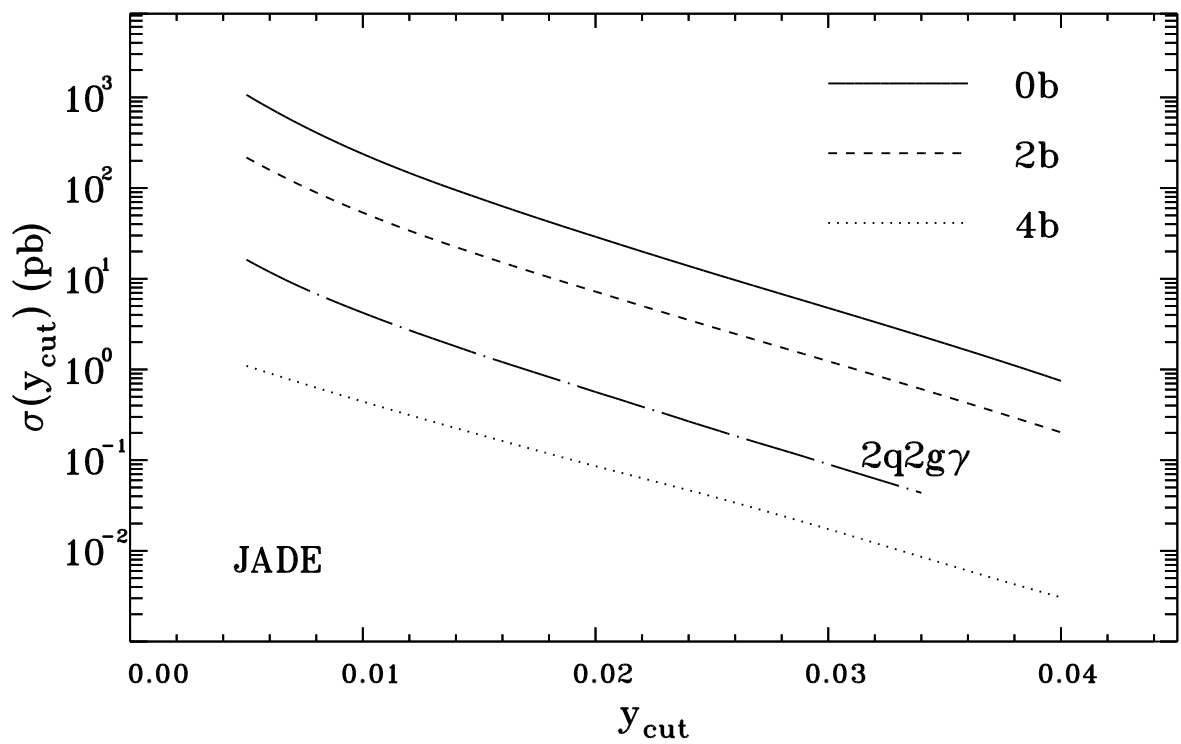
\item[Fig. 3] Total cross section  as a
function of $y_{cut}$ in the JADE and in the DURHAM scheme
for four $b$--quarks plus one gluon--jet production (dotted line),
for the production of
two $b$--quarks plus any combination of three light--parton jets
(dashed line),
and for the sum of all contributions to fiv--jet production including
only light partons (continuous line).

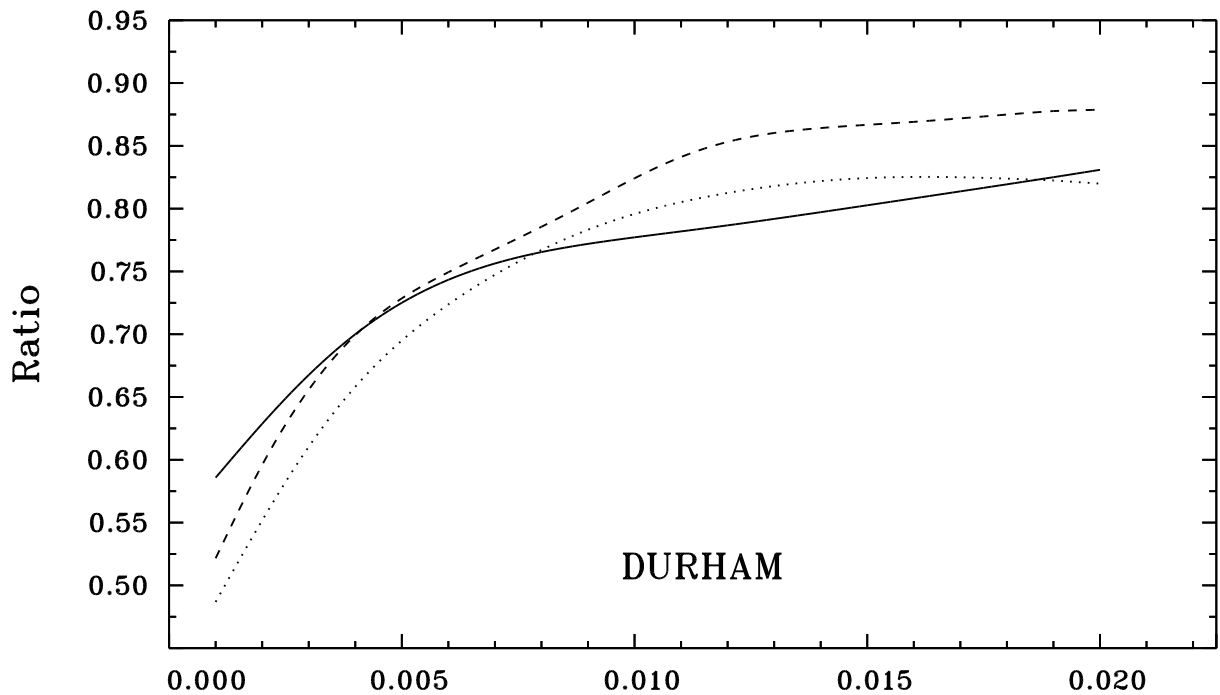
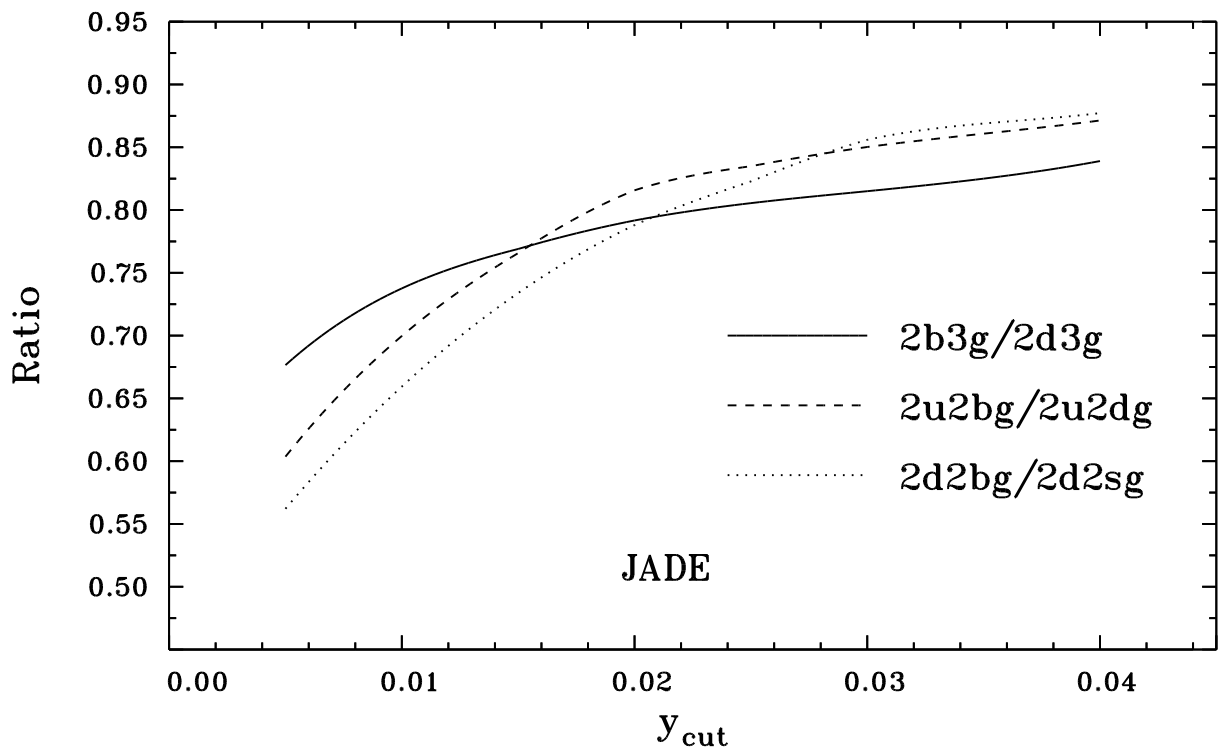
\item[Fig. 4] Ratio of massive to massless cross sections for
  $\sigma(2b3g)/\sigma(2d3g)$
(continuous line), $\sigma(2u2bg)/\sigma(2u2dg)$ (dashed line)
and $\sigma(2d2bg)/\sigma(2d2sg)$ (dotted line) in the JADE and DURHAM
recombination schemes as a function of $y_{cut}$.

\end{description}

\newpage

{\renewcommand {\arraystretch}{1.5}
\begin{tabular}{|c|c|}
\hline
$\lambda_1\lambda_2\lambda_3\lambda_4$ &
%% FOLLOWING LINE CANNOT BE BROKEN BEFORE 80 CHAR
$Z(p_1,\lambda_1;p_2,\lambda_2;p_3,\lambda_3;p_4,\lambda_4;c_R,c_L;c'_R,c'_L)$\\
\hline \hline
$++++$ & $-2[s(3,1)\,t(4,2)c'_Rc_R
-\mu_1\mu_2\eta_3\eta_4c'_Rc_L
-\eta_1\eta_2\mu_3\mu_4c'_Lc_R]$  \\ \hline
$+++-$ & $-2\eta_2c_R[s(4,1)\mu_3c'_L-s(3,1)\mu_4c'_R]$ \\ \hline
$++-+$ & $-2\eta_1c_R[t(2,3)\mu_4c'_L-t(2,4)\mu_3c'_R]$ \\ \hline
$+-++$ & $-2\eta_4c'_R[s(3,1)\mu_2c_R-s(3,2)\mu_1c_L]$ \\ \hline
$++--$ & $-2[s(1,4)\,t(2,3)c'_Lc_R
-\mu_1\mu_2\eta_3\eta_4c'_Lc_L
-\eta_1\eta_2\mu_3\mu_4c'_Rc_R]$  \\ \hline
$+-+-$ & $0$ \\ \hline
$+--+$ & $-2[\mu_1\mu_4\eta_2\eta_3c'_Lc_L
+\mu_2\mu_3\eta_1\eta_4c'_Rc_R
-\mu_2\mu_4\eta_1\eta_3c'_Lc_R
-\mu_1\mu_3\eta_2\eta_4c'_Rc_L]$ \\ \hline
$+---$ & $-2\eta_3c'_L[s(2,4)\mu_1c_L-s(1,4)\mu_2c_R]$ \\ \hline
\end{tabular}
\begin{center}
Table I
\end{center}

\vspace{1. in}

\begin{center}
\begin{tabular}{|c|c|}
\hline
$\lambda_1\lambda_3$ &
$X(p_1,\lambda_1;p_2;p_3,\lambda_3)$ \\ \hline\hline
$++$  & $(\mu_1 \eta_2 + \mu_2\eta_1)
         (\mu_3\eta_2 + \mu_2\eta_3 ) + s(1,2)\, t(2,3)$ \\ \hline
$+-$ & $(\mu_1\eta_2 + \mu_2\eta_1)   s(2,3)
      +  (\mu_2\eta_3 + \mu_3\eta_2 )s(1,2)$ \\ \hline
\end{tabular}
\end{center}
\begin{center}
Table II
\end{center}

\vspace{1. in}

\begin{center}
\begin{tabular}{|c||c|c|c|c|c|c|}
\hline
     & 345 & 453 & 534 & 543 & 354 & 435 \\ \hline \hline
 345 & 64  &  1  &  1  & 10  & -8  & -8  \\ \hline
 453 &  1  & 64  &  1  & -8  & 10  & -8  \\ \hline
 534 &  1  &  1  & 64  & -8  & -8  & 10  \\ \hline
 543 & 10  & -8  & -8  & 64  &  1  &  1  \\ \hline
 354 & -8  & 10  & -8  &  1  & 64  &  1  \\ \hline
 435 & -8  & -8  & 10  &  1  &  1  & 64  \\ \hline
\end{tabular}
\end{center}
\begin{center}
Table III
\end{center}

}

\newpage

{\renewcommand {\arraystretch}{1.5}
\begin{center}
\begin{tabular}{|c||c|c|c|c|c|}
\hline
     &$A_1$&$A_2$&$A_3$&$A_4$&$A_5$\\ \hline \hline
$A_1$& 24  & -3  &  -6 & 21  & -27 \\ \hline
$A_2$& -3  & 24  &  21 & -6  & 27  \\ \hline
$A_3$& -6  & 21  &  24 & -3  & 27  \\ \hline
$A_4$& 21  & -6  &  -3 & 24  & -27 \\ \hline
$A_5$& -27 & 27  &  27 & -27 & 54  \\ \hline
\end{tabular}
\end{center}
\begin{center}
Table IV
\end{center}

\vspace{1. in}

\begin{center}
\begin{tabular}{|c||c|c|c|c|c|}
\hline
     &$A_1$&$A_2$&$A_3$&$A_4$&$A_5$\\ \hline \hline
$B_1$& -8  &  1  &  10 &  1  &  9  \\ \hline
$B_2$&  1  & 10  &  1  & -8  &  9  \\ \hline
$B_3$& 10  &  1  &  -8 &  1  &  -9 \\ \hline
$B_4$&  1  & -8  &  1  & 10  &  -9 \\ \hline
$B_5$&  9  &  9  &  -9 &  -9 &  0  \\ \hline
\end{tabular}
\end{center}
\begin{center}
Table V
\end{center}

}

\end{document}